\documentclass[lettersize,journal]{IEEEtran}

\usepackage{amsmath,amsfonts}
\usepackage{algorithmic}
\usepackage{algorithm}
\usepackage{array}
\usepackage[caption=false,font=normalsize,labelfont=sf,textfont=sf]{subfig}
\usepackage{textcomp}
\usepackage{stfloats}
\usepackage{url}
\usepackage{verbatim}
\usepackage{graphicx}
\usepackage{color}

\usepackage{bm}

\usepackage{acro}
\usepackage{tabularray}
\acsetup{single}

\RenewAcroTemplate[list]{description}{%
	\acroheading
	\acropreamble
	\begin{LaTeXdescription}
		\acronymsmapF{%
			\item[\acrowrite{short}\acroifT{alt}{/\acrowrite{alt}}]
			\acrowrite{list}%
			\acroifanyT{foreign,extra}{ (}%
			\acroifT{foreign}{\acrowrite{foreign}\acroifT{extra}{, }}%
			\acroifT{extra}{\acrowrite{extra}}%
			\acroifanyT{foreign,extra}{)}%
			\acropagefill
			\acropages
			{\acrotranslate{page}\nobreakspace}
			{\acrotranslate{pages}\nobreakspace}%
		}
		{\item\AcroRerun}
	\end{LaTeXdescription}
}

\usepackage[
	style=ieee,
	dashed=false,
	maxbibnames=20,
	minbibnames=2,
	maxcitenames=1,
	mincitenames=1,
	backend=biber,
	defernumbers=false,
	sorting=none,
	useprefix=false
	]{biblatex}
	
\usepackage{tikz}
\usetikzlibrary{arrows,shapes,intersections,calc,angles,decorations.pathreplacing,intersections,quotes,angles,positioning,fit,petri,through,shadows,plotmarks,spy}
\newlength{\unit}
	
\DeclareAcronym{2D}{
	short=2-D,
	long=2-dimensional,
}

\DeclareAcronym{3D}{
	short=3-D,
	long=3-dimensional,
}

\DeclareAcronym{4D}{
	short=4-D,
	long=4-dimensional,
}

\DeclareAcronym{EM}{
	short=EM,
	long=expectation-maximization,
}	

\DeclareAcronym{OSEM}{
	short=OSEM,
	long=ordered-subset EM,
}

\DeclareAcronym{ADMM}{
	short=ADMM, 
	long=alternating direction method of multipliers,
}

\DeclareAcronym{SUV}{
	short=SUV, 
	long=standardized uptake value,
}

\DeclareAcronym{PET}{
	short=PET, 
	long=positron emission tomography,
}

\DeclareAcronym{SPECT}{
    short=SPECT,
    long= single-photon emission computed tomography,
}

\DeclareAcronym{LOR}{
	short=LOR,
	long=line of response,
	long-plural-form = lines of response,
}

\DeclareAcronym{PML}{
	short=PML,
	long=penalized maximum log-likelihood,
}

\DeclareAcronym{MRI}{
	short=MRI, 
	long=MR imaging,
}

\DeclareAcronym{MR}{
	short=MR, 
	long=magnetic resonance,
}

\DeclareAcronym{PMT}{
	short=PMT, 
	long=photomultiplier tube,
}

\DeclareAcronym{MSA}{
	short=MSA, 
	long=multi-head self-attention,
}

\DeclareAcronym{FBP}{
	short=FBP,
	long=filtered-backprojection,
}

\DeclareAcronym{ET}{
	short=ET, 
	long= emission tomography,
}

\DeclareAcronym{CT}{
	short=CT, 
	long=computed tomography,
}

\DeclareAcronym{MBIR}{
	short=MBIR, 
	long=model-based iterative reconstruction,
}

\DeclareAcronym{LBFGS}{
	short=L-BFGS,
	long=limited-memory  Broyden-Fletcher-Goldfarb-Shanno,
}

\DeclareAcronym{XCAT}{
	short=XCAT,
	long=extended cardiac-torso,
}

\DeclareAcronym{AI}{
	short=AI, 
	long=artificial intelligence,
}

\DeclareAcronym{NN}{
	short=NN, 
	long=neural network,
}

\DeclareAcronym{ResNet}{
	short=ResNet, 
	long=residual neural network,
}

\DeclareAcronym{N2N}{
	short=N2N, 
	long=Noise2Noise,
}

\DeclareAcronym{CNN}{
	short=CNN, 
	long=convolutional neural network,
}

\DeclareAcronym{FCN}{
	short=FCN, 
	long=fully convolutional network,
}

\DeclareAcronym{GAN}{
	short=GAN, 
	long=generative adversarial network,
}

\DeclareAcronym{CycleGAN}{
	short=CycleGAN, 
	long=cycle-consistent GAN,
}

\DeclareAcronym{cGAN}{
	short=cGAN, 
	long=conditional GAN,
}

\DeclareAcronym{WGAN}{
	short=W-GAN, 
	long=Wasserstein GAN,
}

\DeclareAcronym{LA-GAN}{
	short=LA-GAN, 
	long=locality adaptive multi-modality GAN,
}

\DeclareAcronym{CNCL}{
	short=CNCL, 
	long=content-noise complementary learning,
}

\DeclareAcronym{ViT}{
	short=ViT, 
	long=vision transformer,
}

\DeclareAcronym{WT}{
	short=WT, 
	long=wavelet transform,
}

\DeclareAcronym{VAE}{
	short=VAE, 
	long=variational autoencoder,
}
\DeclareAcronym{DAE}{
	short=DAE, 
	long=deep denoising autoencoder,
}

\DeclareAcronym{CAE}{
	short=CAE, 
	long=convolutional autoencoder,
}

\DeclareAcronym{DIP}{
	short=DIP, 
	long=deep image prior,
}

\DeclareAcronym{DDIP}{
	short=DDIP, 
	long=double DIP,
}

\DeclareAcronym{MSE}{
	short=MSE, 
	long=mean squared error,
}
\DeclareAcronym{RMSE}{
	short=RMSE, 
	long=root MSE,
}
\DeclareAcronym{SSIM}{
	short=SSIM, 
	long=structural similarity index measure,
}

\DeclareAcronym{RED}{
	short=RED, 
	long=regularization by denoising,
}

\DeclareAcronym{NLM}{
	short=NLM, 
	long= nonlocal means ,
}

\DeclareAcronym{PVE}{
	short=PVE, 
	long= partial volume effect,
}

\DeclareAcronym{PVC}{
	short=PVC, 
	long= partial volume correction,
}

\DeclareAcronym{PSF}{
	short=PSF, 
	long= point spread function,
}

\DeclareAcronym{OOD}{
	short=OOD, 
	long=out-of-distribution,
}

\DeclareAcronym{MP}{
	short=MP, 
	long=myocardial perfusion,
}

\DeclareAcronym{SA}{
	short=SA, 
	long=self-attention,
}

\DeclareAcronym{PSNR}{
	short=PSNR, 
	long=peak signal-to-noise ratio,
}

\DeclareAcronym{SSL}{
	short=SSL, 
	long=self-supervised learning,
}

\DeclareAcronym{DM}{
	short=DM, 
	long=diffusion  model,
}

\DeclareAcronym{CDM}{
	short=CDM, 
	long=conditional diffusion  model,
}

\DeclareAcronym{DDPM}{
	short=DDPM, 
	long=denoising diffusion probabilistic model,
}

\DeclareAcronym{PDF}{
	short=PDF, 
	long=probability distribution function,
}

\DeclareAcronym{WB}{
	short=WB, 
	long=whole-body,
}

\newcommand{\boldr}{\bm{r}}

\newcommand{\boldx}{\bm{x}}
\newcommand{\boldy}{\bm{y}}
\newcommand{\boldz}{\bm{z}}

\newcommand{\boldmu}{\bm{\mu}}

\newcommand{\boldF}{\bm{F}}

\newcommand{\boldP}{\bm{P}}

\newcommand{\boldybar}{\bar{\bm{y}}}
\newcommand{\ybar}{\bar{y}}

\newcommand{\R}{\mathbb{R}}

\newcommand{\transp}{^\top}
\newcommand{\argmin}{\operatornamewithlimits{arg\,min}}

\title{A Review on Low-Dose Emission Tomography Post-Reconstruction Denoising with Neural Network Approaches}

\author{Alexandre Bousse,~\IEEEmembership{Member,~IEEE}, Venkata Sai Sundar Kandarpa,  Kuangyu Shi,~\IEEEmembership{Member,~IEEE}, Kuang Gong,~\IEEEmembership{Member,~IEEE}, Jae Sung Lee,\IEEEmembership{Senior Member,~IEEE}, Chi Liu,~\IEEEmembership{Senior Member,~IEEE},  Dimitris Visvikis,~\IEEEmembership{Fellow,~IEEE} 
	
	\thanks{This work did not involve human subjects or animals in its research.}
	\thanks{This work was supported  by the French National Research Agency (ANR) under grant No ANR-20-CE45-0020, the National Institutes of Health (NIH) under grant No R01EB025468, the Germaine de Stael Program and the Swiss National Science Foundation under grant number 188350}
    \thanks{A. Bousse and V. S. S. Kandarpa equally contributed to this work.}
	\thanks{A. Bousse, V. S. S. Kandarpa  and D. Visvikis are  with Univ. Brest,   LATIM, INSERM UMR 1101, 29238 Brest, France.}
	\thanks{K. Shi is with Lab for Artificial Intelligence \& Translational Theranostics, Dept. Nuclear Medicine, Inselspital, University of Bern, 3010 Bern, Switzerland.}
    \thanks{K. Gong is with The Center for Advanced Medical Computing and Analysis, Massachusetts General Hospital/Harvard Medical School, Boston, MA 02114, USA}
    \thanks{C. Liu is with Department of Radiology and Biomedical Imaging, Yale University, New Haven, CT, USA}
    \thanks{J. S. Lee is with Department of Nuclear Medicine, Seoul National University College of Medicine, Seoul 03080, Korea}

	\thanks{Corresponding authors: A. Bousse, \texttt{bousse@univ-brest.fr}}
	
}

\begin{document}
	
\maketitle

\begin{abstract}
    Low-dose \ac{ET} plays a crucial role in medical imaging, enabling the acquisition of functional information for various biological processes while minimizing the patient dose. However, the inherent randomness in the photon counting process is a source of noise which is amplified in low-dose \ac{ET}. This review article provides an overview of existing post-processing techniques, with an emphasis on deep \ac{NN} approaches. Furthermore, we explore future directions in the field of \ac{NN}-based low-dose \ac{ET}. This comprehensive examination sheds light on the potential of deep learning in enhancing the quality and resolution of low-dose \ac{ET} images, ultimately advancing the field of medical imaging.
\end{abstract}


\printacronyms[]

\section{Introduction}

\IEEEPARstart{T}{he} main components of \ac{ET} are \ac{PET} and \ac{SPECT}. They measure the radio-tracer distribution administered to the patient via gamma radiation arising from  radioactive decay and have multiple use cases including oncology, cardiology, neurology, etc. The ability to get functional information on the various biological processes distinguish them from other imaging modalities such as \ac{MRI} and \ac{CT}.

Radioactive decay is a random process which entails the difficulty in precise production of the images in \ac{ET}. Noise or the speckled variation in \ac{ET} images is caused by the inherent randomness of the photon counting process. In order to ensure patient safety, research has been extensively conducted in the regime of low-dose \ac{ET} imaging \cite{mattsson2011radiation,gimelli2018strategies}.  The reduction in the dose administered to the patient further adds to the challenge of obtaining a clear image. 

\Ac{ET} images are reconstructed from the measured gamma rays, which is an ill-posed inverse problem subject to noise amplification.  The images suffer from \acp{PVE} due to the low intrinsic resolution of the imaging systems, as well as positron range for \ac{PET}. The resolution  of the reconstructed images can be improved with \ac{MBIR} that incorporate the \ac{PSF} in the system matrix. However, this further contributes to the ill-posedness of the inverse problem, which results in more noise. Hence, a number of post-processing techniques have been proposed in this regard. \Ac{AI} and more specifically deep learning-based methods (i.e., deep \acp{NN}) have been very effective in denoising and super resolution and hence have been explored at length in \ac{ET}.  The focus of this article is to discuss at length the \ac{NN} approaches that have been proposed for image denoising in low-dose \ac{ET}. 

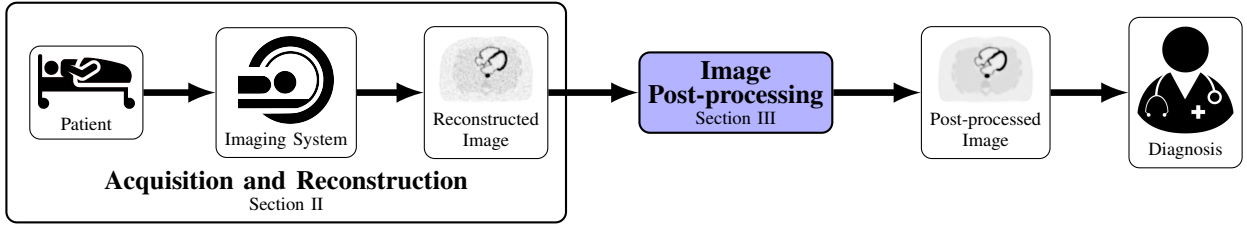
\begin{figure*}
	\centering
	\begin{tikzpicture}
		
		\setlength{\baselineskip}{8pt}

		\node (patient) at (0,0)  [rounded corners, draw, align=center] {
			\includegraphics[width=0.07\linewidth]{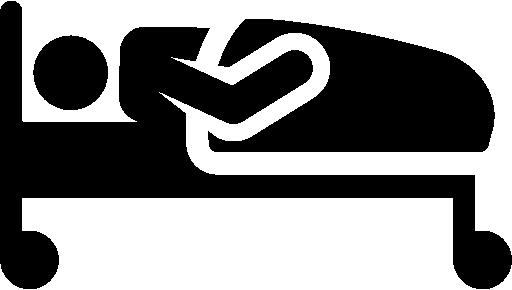} \\ \scriptsize Patient
		} ;
		
		\node (scan) at ([xshift=2.0\unit]patient)  [rounded corners, draw, align=center] {
			\includegraphics[width=0.07\linewidth]{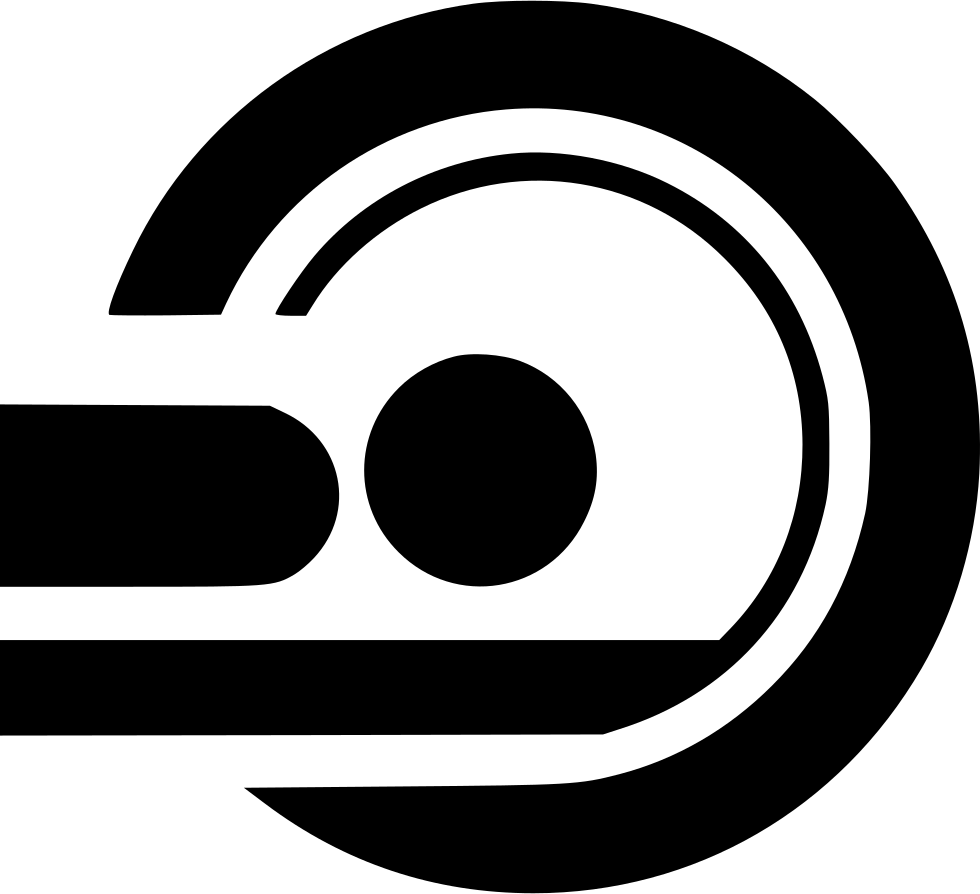} \\ \scriptsize Imaging System
		} ;
		
		\node (scan_and_reco) at ([yshift=-0.2\unit]scan)  [rounded corners, draw, align=center, draw, minimum width=5.6\unit,minimum height=2.2\unit,thick] {  \vspace{6\unit} 
		} ;
		\node (scan_and_reco2) at ([yshift=-0.78\unit]scan_and_reco)  [align=center] {   \textbf{Acquisition and Reconstruction} \\ \scriptsize Section~\ref{sec:recon}
		} ;

		\node (recon) at ([xshift=2.0\unit]scan)  [rounded corners, draw, align=center] {
			\includegraphics[width=0.07\linewidth]{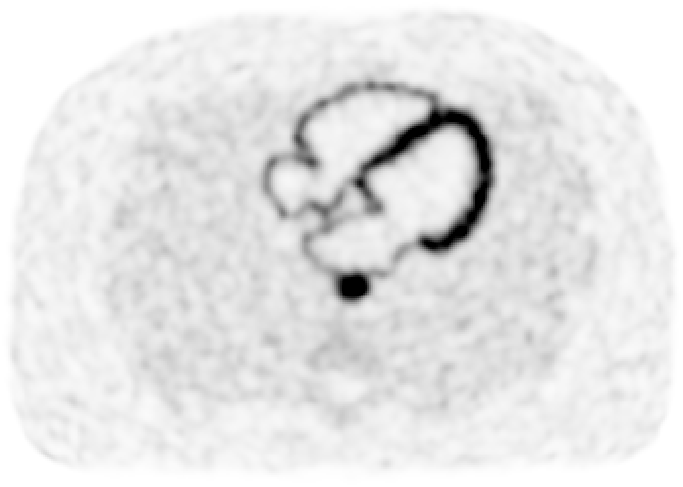} \\ \scriptsize Reconstructed \\ \scriptsize Image
		} ;
		\node (denoising) at ([xshift=2.5\unit]recon)  [align=center,rounded corners, draw, thick, fill=blue!30] { \textbf{Image} \\
			\textbf{Post-processing} \\ \scriptsize Section~\ref{sec:dl}  
		} ;
		\node (denoised) at ([xshift=2.5\unit]denoising)  [rounded corners, draw, align=center] {
			\includegraphics[width=0.07\linewidth]{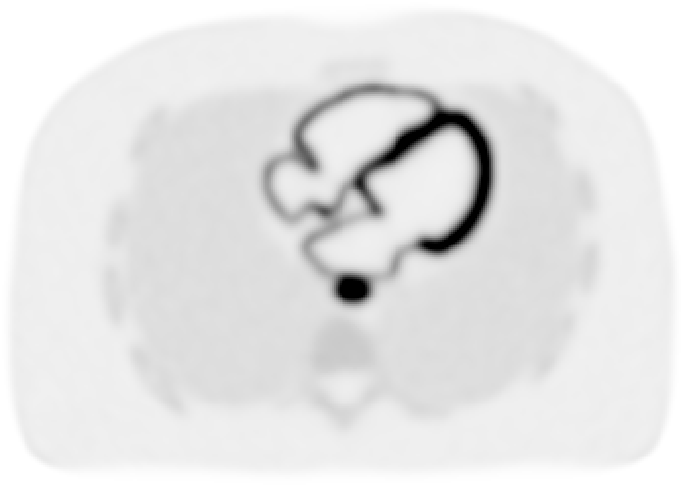} \\ \scriptsize Post-processed \\ \scriptsize Image
		} ;
		\node (diag) at ([xshift=2.0\unit]denoised)  [rounded corners, draw, align=center] {
			\includegraphics[width=0.07\linewidth]{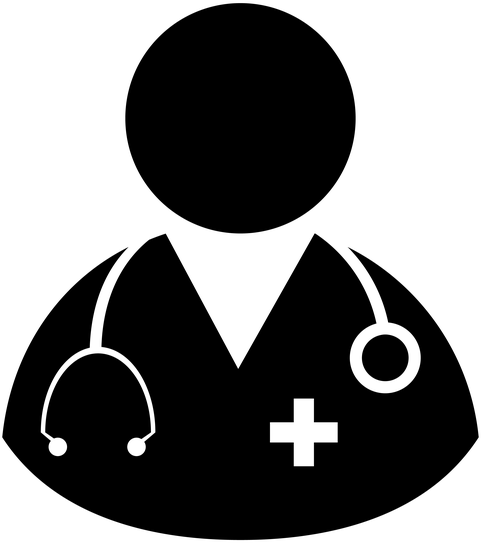} \\ \scriptsize Diagnosis
		} ;

		\draw  [->,>=latex,line width=2.5pt,draw,out=0,in=180] (patient.0) to node [midway,below,align=center](mid1) {} (scan.180) ;	
		\draw  [->,>=latex,line width=2.5pt,draw,out=0,in=180] (scan.0) to node [midway,below,align=center] {} (recon.180) ;
		\draw  [->,>=latex,line width=2.5pt,draw,out=0,in=180] (recon.0) to node []{} (denoising.180) ;	
		\draw  [->,>=latex,line width=2.5pt,draw,out=0,in=180] (denoising.0) to node [] {} (denoised.180) ;	
		\draw  [->,>=latex,line width=2.5pt,draw,out=0,in=180] (denoised.0) to node [] {} (diag.180) ;

	\end{tikzpicture}	

	\caption{Main steps in clinical \ac{ET}. The patient is initially administered with a radioactive tracer. The emission is captured by an imaging system. The mapping of the raw detector data to an image is performed by an image reconstruction method.  Finally, after post-processing, the image is used for diagnosis.}\label{fig:steps_ET}
\end{figure*}

The main steps in the clinical application of \ac{ET}, are highlighted in Fig.~\ref{fig:steps_ET}. The first step is the production of the radio-pharmaceutical followed by its administration to the patient. Following this the emission data is acquired by a detector setup. Using the physical forward model, image reconstruction methods are utilized to map the raw detector data to an image. These steps are summarized in Section~\ref{sec:recon}. Following reconstruction, the image can be post-processed to reduce the noise and improve resolution. Section~\ref{sec:dl} gives an overview of existing post-processing techniques for low-dose \ac{ET} image post-processing, with an emphasis on \ac{NN}-based approaches, and is the main contribution of this paper. Finally, Section~\ref{ref:disc} covers future directions of \ac{NN}-based low-dose \ac{ET}.

\section{Data Acquisition and Image Reconstruction}\label{sec:recon}

\subsection{Data Acquisition}

\Ac{SPECT} imaging is based on the emission of a single gamma photon per each radioactive decay event. These gamma photons are usually collected by a  gamma camera \cite{anger1958scintillation}, which typically consists of a collimator that selects relevant gamma photons to be detected; the gamma photons are  converted into light in the visible spectra by the scintillation crystal. The optical-wavelength photons are sent to a \ac{PMT} that converts light into electrons, generating a detectable current. This current is then measured by an electronics setup, noting the occurrence of the event.
The relative spatial coordinates of the event  are determined by measurements from the point of contact in the \acp{PMT}. These events are stored in histograms based on their position, resulting in discrete (or vectorized) projection measurement data which is then utilized for image reconstruction.   

\Ac{PET} on the other hand uses positron emitting radioisotope. The positron interacts with an electron resulting in an annihilation event  that produces two gamma photons moving in nearly opposite direction. These gamma photons are simultaneously detected (coincidence event) by circularly arranged detector elements. Due to this inherent feature of \ac{PET}, collimators are not present in the detector system. The coordinates of each decay event are recorded, through the detection of the corresponding coinciding pair of gamma photons. The binning of these detected coincidence events results in the projection data, used for reconstruction.

\ac{ET} image reconstruction and post-processing requires a formalism that we briefly describe here. The radiotracer distribution takes the form of an image vector $\boldx = [x_1,\dots, x_m]\transp \in \R^m$, `$\transp$' denoting the matrix transpositioon, where each entry $x_j$ denotes the radiotracer concentration at voxel $j$ (in Bq per voxel). In both \ac{SPECT} and \ac{PET}, the imaging system is modeled by a system matrix $\boldP \in \R^{n\times m}$ where for all $(i,j)$ each entry $[\boldP]_{i,j}$ is the probability that an emission in voxel $j$ is detected along the $i$th \ac{LOR}, taking into account the geometry of the system, the linear attenuation, the sensitivity of the detectors and the intrinsic resolution. For each \ac{LOR} $i=1,\dots,n$, the expected number of detections given a radiotracer distribution $\boldx$ is
\begin{equation}\label{eq:expectation}
	\ybar_i(\boldx) = \tau[\boldP\boldx]_i + r_i
\end{equation} 
where $\tau$ is the scan duration and $r_i$ is a background term comprising expected scatter as well as randoms (for \ac{PET}), and the number of detection is a random variable $y_i$ that follows a Poisson distribution centered in $\boldybar(\boldx)$, i.e.,
\begin{equation}\label{eq:poisson}
	y_i \sim \mathrm{Poisson}\left(\ybar_i(\boldx)\right) \, .
\end{equation}  
In the following we denote $\boldy = [y_1,\dots,y_n]\transp$ and $\boldybar(\boldx) = [\ybar_1(\boldx),\dots,\ybar_n(\boldx)]\transp$ the measured and expected data respectively, and $\boldr = [r_1,\dots,r_n]\transp$ the background events vector.

\subsection{Reconstruction}

Reconstructing an image $\boldx^\mathrm{rec}$ corresponds to solving the following inverse problem:
\begin{equation}\label{eq:inv_problem}
	\text{finding $\boldx^\mathrm{rec}$ s.t.}\quad \boldy \approx \boldybar(\boldx^\mathrm{rec}) \, ,
\end{equation}
This can be achieved by analytical inversion of $\boldP$ applied to $\frac{1}{\tau}(\boldy - \boldr)$, also known as \ac{FBP} (see for example \cite{Natterer2001mathematics}). Unfortunately, the inverse problem~\eqref{eq:inv_problem} is ill-posed and direct inversion leads to noise amplification which is impractical for low-dose imaging. Moreover, the inversion relies on an idealized model that does not incorporate resolution modeling. Finally, solving  \eqref{eq:inv_problem} does not guarantee positivity of the solution.

Another approach consists in finding an estimate $\boldx^\mathrm{rec}$ by \ac{PML}
\begin{equation}\label{eq:pml}
	   \boldx^\mathrm{rec} \in	 \argmin_{\boldx \in \R^m_+} \, \ell\left(\boldy , \boldybar(\boldx)\right) + \beta R(\boldx)\end{equation} 
where $\ell\left(\boldy , \boldybar\right)$ is the Poisson negative log-likelihood of the expectation $\boldybar$ given the measurement $\boldy$, $R$ is a penalty, or \emph{prior}, that enforces image smoothness and  $\beta>0$ is a weight. Solving \eqref{eq:pml} can be achieved by \ac{MBIR} such as the \ac{EM} algorithm \cite{shepp1982maximum} or \ac{OSEM}  algorithm \cite{Hudson1994} in absence of penalty (i.e., $\beta=0$), and modified \ac{EM} \cite{depierro1995modified} with a smooth convex penalty. Reconstructing the image $\boldx$ by \ac{PML} \eqref{eq:pml} reduces the noise as compared with analytical inversion \eqref{eq:inv_problem} by (i) the incorporation of the stochastic model of the noise in $L$ and (ii) by the presence of the penalty that controls the noise. Typical penalty term $R$ includes quadratic smoothness penalty \cite{qi1998fully}, the edge-preserving Huber penalty \cite{huber2004robust} or the relative difference prior \cite{nuyts2002concave}. Anatomical priors have been used to smooth the \ac{PET} image while preserving image resolution by taking advantage of high-resolution  anatomical images such as \ac{CT} or \ac{MRI} \cite{bowsher2004utilizing,somayajula2010pet,kazantsev2012anatomically,pedemonte20114,ehrhardt2016pet}. More recently,  \citeauthor{sudarshan2018joint}~\cite{sudarshan2018joint,sudarshan2020joint}  introduced patch-based dictionary learning for joint \ac{PET}/\ac{MR} image reconstruction. 

\Ac{PML} methods have played a pivotal role in \ac{ET} image reconstruction. These methods have proven effective in managing noise levels and improving image quality. However, in the context of low-dose \ac{ET} imaging, striking a suitable balance between noise reduction and preservation of essential image details necessitates careful tuning of the prior weight. Failure to do so may lead to undesired over-smoothing artifacts. Furthermore, anatomically-guided penalties, while valuable in certain cases, can introduce artifacts when there is misalignment between the activity and anatomical images. It is also worth noting that the applicability of \ac{PML} techniques is contingent upon the availability of raw data, which may not always be obtainable.

\section{Deep Learning-based Image Post-processing}\label{sec:dl}

When \ac{ET} raw data are not available, images can be post-processed to improve their quality. Prior to the advent of deep learning, conventional image post-processing methods were employed for incorporating corrections in  \ac{ET} images. The first image post-processing task is to improve the image resolution, namely, \acf{PVC}. \Ac{PVC} first achieved with deconvolution techniques, which consists of correcting  for the image \ac{PSF} using iterative techniques, such as the Van~Citert algorithm~\cite{vancitert1931einfluss} or the Richardson-Lucy algorithm~\cite{richardson1972bayesian,lucy1974iterative}. However, deconvolution is an ill-posed inverse problem and leads to noise amplification, which is non-practical for low-dose imaging. Therefore, it is necessary to deploy adequate techniques to control the noise. This topic has been the subject to numerous works over the last decade \cite{rousset2007partial,erlandsson2012review,hutton2013approach,erlandsson2016mr,thomas2016petpvc}.

The inherent relationship between resolution and noise poses a significant challenge. Enhancing image resolution typically leads to increased noise levels, while reducing noise tends to compromise resolution. Addressing both aspects simultaneously requires a paradigm shift towards \ac{AI}-based approaches.  \Acp{NN} have revolutionized imaging ever since the performance of AlexNet in the ImageNet challenge \cite{krizhevsky2017imagenet}. In medical imaging, they have contributed to image segmentation, cancer detection, registration, reconstruction, etc. \cite{litjens2017survey}. In the context of low-dose \ac{ET} imaging, they have been widely implemented to bring about improvements in images reconstructed by traditional algorithms.

This section reviews the \ac{NN}-based low-dose \ac{ET} denoising methods and categorizing them based on their \ac{NN} design. We have broadly categorized the methods into supervised methods, \ac{SA} mechanisms, unsupervised methods, multi-modality (i.e., additional anatomical information from another modality) and  \acp{DM}. These subsections have further been divided into subcategories to distinguish and highlight specific workings of the approaches. Note that, although \ac{SA} mechanisms  are typically categorized as supervised methods, recognizing their increase in popularity and effectiveness, we have chosen to dedicate a separate section to these mechanisms, with an  emphasis on transformers.

\subsection{Supervised Methods}\label{sec:supervised}

Machine learning methods that require labeled data for training come under this category. Owing to the data revolution and the availability of annotated datasets for different tasks along with competitions, the most popular methods for denoising are supervised. 

A reconstructed image   $\boldx^\mathrm{rec}$  produced by traditional methods like \ac{FBP} (by solving \eqref{eq:inv_problem}) or \ac{PML} (by solving \eqref{eq:pml})  is processed through an image-to-image \ac{NN} $\bm{F}_{\bm{\theta}}$ depending of a trained parameter $\bm{\theta}^\star$ as
\begin{equation}\label{eq:post_pro}
	\boldx^{\mathrm{pp}} = \bm{F}_{\bm{\theta}^\star} (\boldx^\mathrm{rec})
\end{equation} 
where $\boldx^{\mathrm{pp}}$ is the post-processed image. Supervised training of $\bm{\theta}^\star$ is generally achieved using a training dataset of $K$ noisy/clean image pairs $(\boldx^{\mathrm{noisy}}_k,\boldx^{\mathrm{clean}}_k)$, $k=1,\dots,K$, as 
\begin{equation}\label{eq:sup_training}
	\bm{\theta}^\star = \argmin_{\bm{\theta}} \,  \sum_{k=1}^K L\left(\bm{F}_{\bm{\theta}} \left(\boldx^\mathrm{noisy}_k\right), \boldx^{\mathrm{clean}}_k\right)
\end{equation}
where $L(\cdot,\cdot)$ is a loss function.

There are many possible variations of the \ac{NN} $\bm{F}_{\bm{\theta}}$ and its training. This subsection will cover \acp{FCN} and \acp{GAN}.

\subsubsection{Fully Convolutional Networks}

\begin{figure*}
    \centering
    \includegraphics[width=1.0\linewidth]{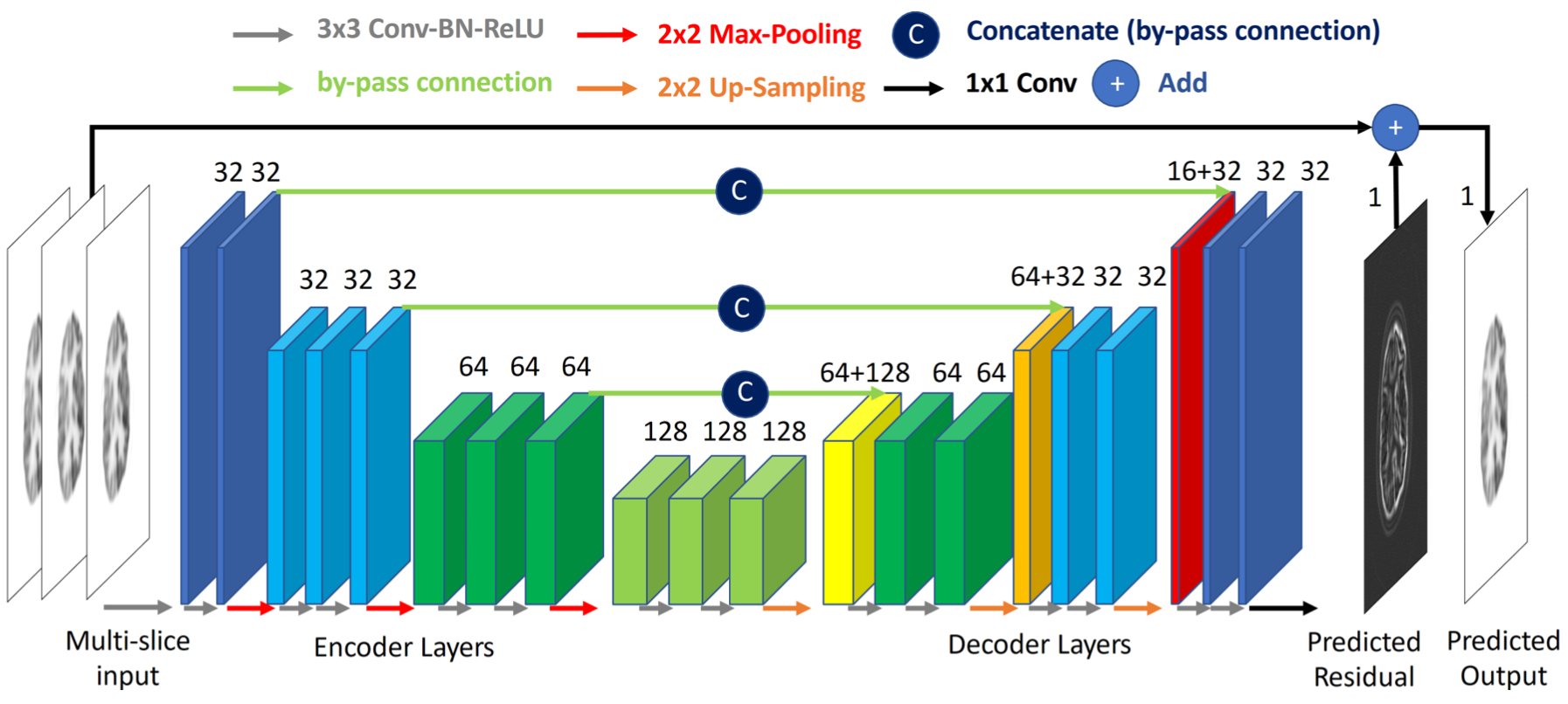}
    \caption{Representation of supervised deep learning-based method from \citeauthor{xu2017200x}~\cite{xu2017200x}.}
    \label{fig:200x}
\end{figure*}

Computer vision tasks like image segmentation, super-resolution, image enhancement, etc., typically utilize \acp{FCN} that are based on \ac{ResNet} \cite{he2016deep} or U-Net \cite{ronneberger2015u}. One of the earliest works that used \acp{FCN} for denoising in \ac{PET} was proposed in \cite{xu2017200x}. The proposed method predicted full-dose \ac{PET} images from \ac{PET} images with dosage reduced by 200 times. The authors used a convolutional encoder-decoder-styled architecture with three convolutional layers on the encoder and decoder part of the network. The encoder consisted of convolutions and max-pooling layers while the decoder consisted of upsampling through bilinear interpolation. In order to tackle the resolution loss experienced in the encoder-decoder type of structure the authors also employed concatenations between the encoder and decoder similar to a U-Net. In addition, a residual connection from the input to the output image was used. This enabled the network to learn the difference between the full-dose and low-dose \ac{PET} images. The network was also trained with multi-slice input so as to help the network distinguish between noise and finer structural details. The loss function of choice in this work was the $L^1$-norm. This approach is shown in Fig.~\ref{fig:200x}. \citeauthor{gong2018pet}~\cite{gong2018pet} proposed an architecture based on \ac{ResNet} consisting of five residual blocks. Owing to the limited amount of real data, the authors initially trained the network on simulated data created using BrainWeb \cite{cocosco1997brainweb} and the \ac{XCAT} phantom \cite{segars2008realistic}. The network was then fine-tuned on real data. Another aspect of this work is the use of perceptual loss to further improve the quality of predicted images. The authors found their proposed method to perform better than traditional Gaussian filtering. Dilated convolution \cite{chen2017deeplab} replaced the convolution operation in the U-Net in the work proposed by \citeauthor{spuhler2020full}~\cite{spuhler2020full}. The network, called dNet, outperformed U-Net for \ac{PET} image denoising. The advantage of dilated convolutions, which were first introduced for image segmentation, is that they remove the requirement of pooling and upsampling operators. The dataset used in this study was from a psychiatric study consisting of 35 patients.  \Ac{DAE} was proposed in \cite{klyuzhin2019use} for dynamic \ac{PET} denoising. The \ac{DAE} was trained on noisy and noiseless spatiotemporal patches of simulated images. Although a promising voxel-level denoising method was proposed, the proposed \ac{DAE} struggled to generalize to test data different from the training data. A \ac{3D} version of the U-Net was used in \cite{schaefferkoetter2020convolutional} for mapping from noisy  $64 \times 64 \times 64$ patches   to noise-reduced patches of the same dimension. The authors trained the network on a lung cancer real dataset. The method was evaluated by three physicians through lesion detection tasks. The proposed method performed better than Gaussian smoothing, but its improvements were limited when it came to the count levels typically observed in a clinical setting. The effect on noise levels for denoising \ac{PET} images was studied in \cite{liu2022personalized}. A personalized weighting strategy for specific noise levels was proposed through the linear blending of results from different models. The authors trained five \ac{3D} U-Nets each with a different noise level in the training data. Along with these, a separate network with all the varied noise levels also was trained. The one-network-for-all model did not generalize well on the testing data with multiple noise levels. The networks trained on noisier images performed better at denoising but introduced more spatial blurring. The final method fused the deep image prior and \ac{RED} approach to obtain a final denoised image. DeepRED optimization was done using \ac{ADMM} \cite{Boyd2010}.

Low-dose \ac{MP} \ac{SPECT} with deep learning was proposed in \cite{ramon2020improving}, where the authors used a \ac{3D} \ac{CNN}. The \ac{3D} network consisted of autoencoders where the encoder and the decoder parts of the network were stacked with convolutions. The network was trained to map from various levels of dosage ($1/2,1/4,\dots,1/16$) to full-dose images. The authors found their method to perform better than the conventional spatial-post filtering method. The network was trained on real patient data reconstructed with both \ac{FBP} and \ac{OSEM} algorithms. A four-layer U-Net was utilized in \cite{reymann2019u} for \ac{SPECT} denoising. The U-Net in this method was trained on simulated \ac{XCAT} phantoms. Low-dose imaging in \ac{SPECT} through the reduction in acquisition time and projection angles was explored in \cite{shiri2020standard}. The authors used a \ac{ResNet} for mapping from low-dose to full-dose images. The dataset consisted of \ac{MP} \ac{SPECT} images of 363 patients recontructed with \ac{OSEM} algorithm. A study on weighted loss functions for varying levels of statistics based on inter-subject changes   for deep learning-based \ac{SPECT} denoising was proposed in \cite{liu2021accounting}.

\subsubsection{Generative Adversarial Networks}\label{sec:gans}

\Acp{GAN} \cite{goodfellow2014advances} are a special type of \ac{NN} model consisting of two units, with the generator unit synthesizing candidates while the discriminator unit attempts to decipher whether the candidate’s images are synthetic or real. The development of \acp{GAN} has strengthened the capability of \acp{NN} in this regard, allowing them to capture complex probability distributions. \citeauthor{lu2019investigation}~\cite{lu2019investigation} investigated the accuracy of deep learning-based denoising methods including \acp{GAN} for \ac{PET} imaging of small lung nodules, focusing on quantitative accuracy and visual image quality. \citeauthor{ouyang2019ultra}~\cite{ouyang2019ultra} explored \acp{GAN} with feature matching and task-specific perceptual loss in the restoration of  amyloid \ac{PET}. \citeauthor{jeong2021restoration}~\cite{jeong2021restoration} demonstrated that \ac{GAN}-based restoration of amyloid \ac{PET} scans did not affect physician interpretation, indicating that the restored images were consistent with the original scans and preserved their diagnostic value.

\Ac{cGAN} was introduced by \citeauthor{wang20183d}~\cite{wang20183d} to recover full-dose brain [\textsuperscript{18}F]FDG \ac{PET} images from low-dose measurements. \citeauthor{xue2021cross}~\cite{xue2021cross} confirmed the cross-scanner and cross-tracer capability of customized \ac{cGAN}, where the training was done from [\textsuperscript{18}F]FDG \ac{PET} on one scanner and the test was performed on [\textsuperscript{18}F]FET and [\textsuperscript{18}F]Florbetapir \ac{PET} imaging of different scanners. 

\Ac{CycleGAN} was applied to recover full-dose \ac{WB} \ac{PET} from low-dose measurements \cite{lei2019whole}. \citeauthor{zhou2020supervised}~\cite{zhou2020supervised} confirmed that their \ac{CycleGAN} preserves edges and \acp{SUV} from the restored low-dose dataset with biopsy-proven primary lung cancer or suspicious radiological abnormalities. A supervised \ac{GAN} with the cycle-consistency loss, Wasserstein distance loss, and an additional supervised learning loss, named as S-CycleGAN was demonstrated to outperformed \ac{3D}-\ac{cGAN} in the recovery of low-dose brain \ac{PET} \cite{zhao2020study}. The \ac{CycleGAN} has demonstrated the advantage to train non-synthetic low-dose \ac{WB} [\textsuperscript{18}F]FDG \ac{PET} scans together with separate full-dose \ac{WB} [\textsuperscript{18}F]FDG \ac{PET} scans in a study by \citeauthor{sanaat2021deep}~\cite{sanaat2021deep}.

\begin{figure*}
    \centering
    \includegraphics[width=1.0\linewidth]{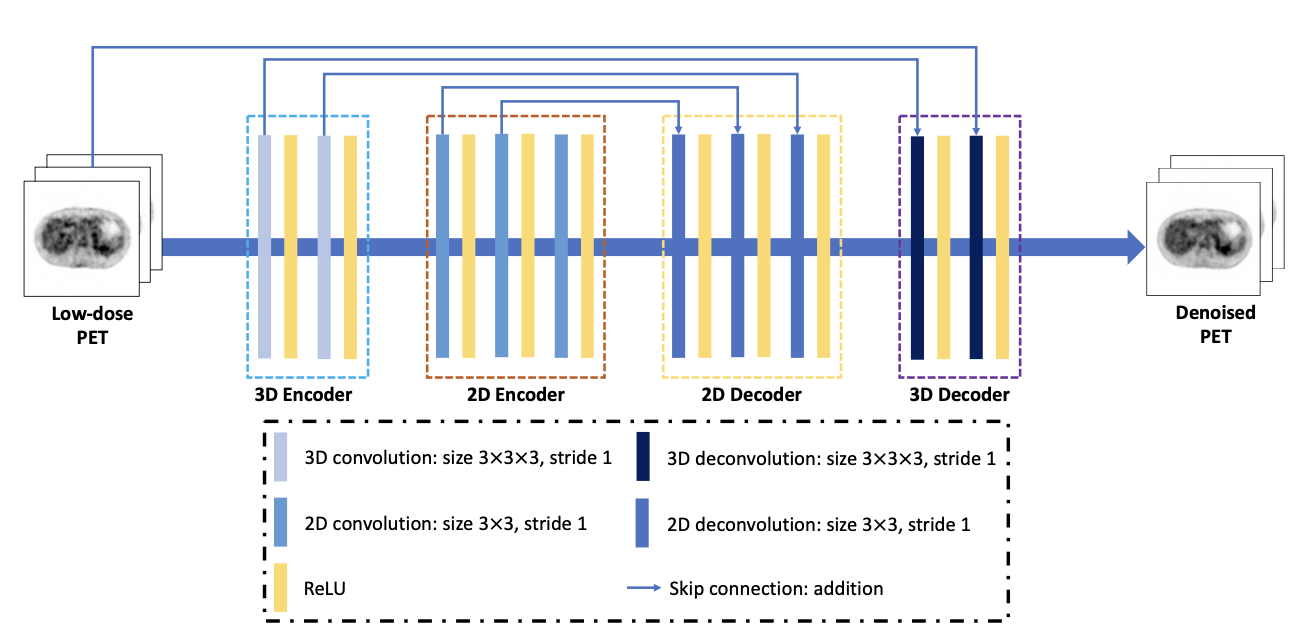}
    \caption{Structure of PT-WGAN from \citeauthor{gong2020parameter}~\cite{gong2020parameter}.}
    \label{fig:PT-WGAN}
\end{figure*}

\citeauthor{gong2020parameter}~\cite{gong2020parameter} proposed a parameter-transferred \ac{WGAN}, namely, PT-GAN,  with a task-specific initialization for low-dose \ac{PET} image denoising without compromising structural details. A representation of PT-GAN is shown in Fig.~\ref{fig:PT-WGAN}. \citeauthor{du2020iterative}~\cite{du2020iterative} developed a cascaded data consistency \ac{GAN}  to recover high-quality \ac{PET} images from \ac{FBP}-reconstructed \ac{PET} images with streaking artifacts and high noise. \citeauthor{geng2021content}~\cite{geng2021content} developed a \ac{CNCL} pipeline using \ac{GAN} to reduce noise in medical images including \ac{CT}, \ac{MR}, and \ac{PET}, which outperformed state-of-the-art denoising algorithms in terms of visual quality, quantitative metrics and robust generalization capability. 

Additional modality such as \ac{MRI} can be added into the training of \ac{GAN}. \citeauthor{wang20183dauto}~\cite{wang20183dauto}  developed a \ac{3D} auto-context-based locality-adaptive multi-modality \ac{GAN} (LA-GAN) for estimating full-dose [\textsuperscript{18}F]FDG \ac{PET} images from low-dose counterparts together with \ac{MR} images and demonstrated the superiority over traditional multi-modality fusion methods for \ac{PET} restoration.  \citeauthor{zhou2021mdpet}~\cite{zhou2021mdpet} introduced a unified motion correction and denoising \ac{GAN} for generating motion-compensated low-noise images from low-dose gated \ac{PET} data. 

\Acp{GAN} can be also applied on denoising directly during the reconstruction from low-count sinogram data. \citeauthor{xue2021lcpr}~\cite{xue2021lcpr} use a LCPR-Net to enforced a cyclic consistency constraint on the least-squares loss of a \ac{GAN} framework to establish a nonlinear end-to-end mapping process from low-count sinograms to full-count \ac{PET} images.  Similarly, an improved \ac{WGAN} framework was employed as a direct \ac{PET} image reconstruction network (DPIR-Net)  to enhance image speed and quality of \ac{PET} reconstruction \cite{hu2020dpir}. 

\Acp{GAN} were also applied on the denoising of \ac{SPECT} images. \citeauthor{sun2022dual}~\cite{sun2022dual} developed a method based on a \ac{3D} \ac{cGAN} for denoising of dual-gated \ac{MP} images. The same group also investigated the denoising performance of \ac{cGAN} in projection-domain and compared it with the denoising in reconstruction-domain for low-dose \ac{MP} \ac{SPECT} imaging \cite{sun2022deep}. \citeauthor{sun2022pix2pix}~\cite{sun2022pix2pix} introduced Pix2Pix \ac{GAN}   for denoising low-dose \ac{MP}  images and found it superior than other denoising methods. They also introduced attention mechanisms in \acp{GAN} for the denoising of fast \ac{MP}  images \cite{sun2023fast}.  \citeauthor{aghakhan2022deep}~\cite{aghakhan2022deep}  developed a \ac{GAN} to predict non-gated standard-dose \ac{SPECT} images in the projection space. Their finding revealed that recovery of underlying signals/information in low-dose images beyond a quarter of the standard dose would not be feasible and adversely affect the clinical interpretation of the resulting images.

\subsection{Self-attention Mechanisms}\label{sec:self-atten}   

The \ac{SA} mechanism was first proposed for natural language processing by  \citeauthor{vaswani2017attention}~\cite{vaswani2017attention}. The generated attention map  is  similar to the weight elements utilized in the \ac{NLM} denoising \cite{chan2013postreconstruction}. This connection was further explained in the work of nonlocal \acp{NN}\cite{wang2018non}. 

During \ac{SA} calculation, features are first extracted from the input to construct the \emph{Query}, \emph{Key}, and \emph{Value} components. The \emph{Query} and \emph{Key} components are then utilized to generate the attention map through a matrix multiplication, scaling, and SoftMax operations. The calculated attention map is multiplied with the \emph{Value} component to obtain the final output. Compared to the widely used convolution operation, the \ac{SA} module has a spatially-variant filter defined by the attention map, which is calculated from the input itself. The \ac{SA} module can be embedded into the popular U-Net and \ac{GAN} structures to further improve the performance as demonstrated in various computer vision tasks \cite{oktay2018attention,zhang2019self}. 

For \ac{PET} image denoising,  \citeauthor{xue20203d}~\cite{xue20203d} proposed embedding the \ac{SA} block into the widely used U-Net structure.  The \ac{SA} can also be embedded into other network structures for \ac{PET} image denoising, such as \ac{GAN} \cite{fu2023aigan} and \ac{CycleGAN} \cite{shang2022short,lei2020low}.  Apart from utilizing only low-dose \ac{PET} images as the input, \citeauthor{onishi2021anatomical}~\cite{onishi2021anatomical} proposed an unsupervised \ac{PET} image denoising framework that incorporated anatomical information into the network architecture via the \ac{SA} mechanism. The attention gates employed in this work aimed to remove \ac{PET} image noise by better utilizing the multi-scale semantic features extracted from the \ac{MR} prior image.  For dynamic \ac{PET} imaging, \citeauthor{li2023deep}~\cite{li2023deep} proposed to directly generate high-quality Patlak images from five-frame sinograms by a network with \ac{SA} blocks, potentially reducing the acquisition time and avoiding the input function needed for parametric imaging. For \ac{PET} image reconstruction, \citeauthor{xie2020generative} ~\cite{xie2020generative} proposed to utilize the U-Net with a \ac{SA} block for image representation and further employ this network representation into the \ac{PET} \ac{MBIR} framework. They also employed additional high-resolution anatomical images as the network input to further improve the reconstruction performance \cite{xie2021anatomically}.

Though \acp{CNN} achieved great success in various medical imaging tasks, the network specifically focused on local spatial information, and the receptive field was also limited. The transformer networks extensively employed \ac{SA} blocks (described in the previous subsection) as the network building blocks, which had the ability to capture long-range information. \Ac{ViT} \cite{dosovitskiy2020image} was the first application of the transformer network to computer vision. The input images were firstly divided into patches, linearly embedded along with the position information, and then fed to the transformer network for image classification. One issue of \ac{ViT} was the quadratically growing computational complexity along the spatial dimension, which limited the receptive field achievable.

\begin{figure*}
    \centering
    \includegraphics[width=1.0\linewidth]{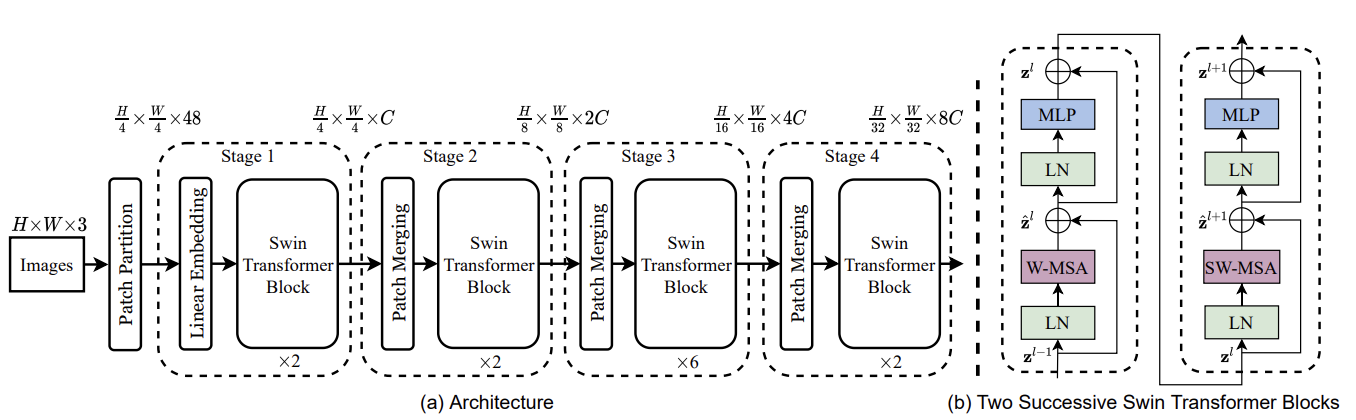}
    \caption{(a) The architecture of the Swin transformer network  and (b) consecutive blocks of the Swin transformer. Reprint from \citeauthor{liu2021swin}~\cite{liu2021swin}.}
    \label{fig:swin}
\end{figure*}

To address this issue,  the Swin Transformer \cite{liu2021swin} was proposed to efficiently calculate local \ac{MSA} using shifted windows that offered linear computational complexity. Compared to \ac{ViT}, the Swin Transformer divided the whole image into windows. The \ac{SA}  was computed based on a shifted windowing scheme, which achieved greater efficiency by limiting \ac{SA} calculation to non-overlapping local windows while also considering cross-window connections. This method is represented in Fig.~\ref{fig:swin}. Instead of \ac{SA} calculation along the spatial domain, Restormer \cite{zamir2022restormer} was recently proposed to efficiently compute global \ac{MSA} along the channel dimension. It had linear computational complexity for image restoration tasks.

The transformer networks were recently applied for \ac{PET} image quality improvement.  \citeauthor{luo20213d}~\cite{luo20213d} proposed a \ac{GAN} embedded with a transformer to perform low-dose \ac{PET} image denoising. The transformer was inserted between the encoder and decoder paths of the generator network, and the training function was based on both voxel-wise estimation error and the adversarial loss.  \citeauthor{jang2022spach}~\cite{jang2022spach} proposed a transformer network that can leverage both spatial and channel information based on local and global \acp{MSA}. Quantitative evaluations based on datasets of different \ac{PET} tracers, i.e., [\textsuperscript{18}F]FDG, [\textsuperscript{18}F]ACBC, [\textsuperscript{18}F]DCFPyL, and [\textsuperscript{68}Ga]DOTATATE, showed that the proposed transformer structure achieved better performance than other reference methods. When utilizing both low-dose \ac{PET} and high-resolution \ac{MR} prior images as the input,  \citeauthor{zhang2022spatial}~\cite{zhang2022spatial} designed a network structure with two paths to extract \ac{PET} and \ac{MR} features and a transformer block to fuse the \ac{PET} and \ac{MR} features. \citeauthor{wang2023low}~\cite{wang2023low} compared five deep learning-based denoising methods, including Swin Transformer and \ac{ViT}, under different \ac{PET} dose levels. Results showed that the Swin Transformer achieved better performance than other reference methods in most evaluation tasks. Apart from \ac{PET} image denoising, \citeauthor{hu2022transem}~\cite{hu2022transem} utilized the transformer network for \ac{PET} image reconstruction following the unrolled-\ac{NN} framework \cite{gong2019mapem,mehranian2020model,lim2020improved}, and the results showed better performance than other unrolled networks where \acp{CNN} were adopted.

\subsection{Unsupervised Methods}\label{sec:unsup}

The advent of unsupervised methods was a result of the limitations that supervised learning methods posed namely large datasets and the effort that goes into annotating them. Two such methods that have been developed in the context of denoising are \ac{DIP} and \ac{N2N} methods. 

\subsubsection{Deep Image Prior}\label{sec:dip}

The \ac{DIP} \cite{ulyanov2018deep} is a learning-free method that demonstrated that a randomly initialized \ac{NN} could be used as a prior for inverse problems like denoising, inpainting, and super-resolution. The \ac{DIP} changes the paradigm of the standard deep denoising approach \eqref{eq:post_pro} in the sense that instead of training the parameter $\bm{\theta}$ from a dataset as described in \eqref{eq:sup_training}, the post-processed image $\boldx^{\mathrm{pp}}$ is obtained from the reconstructed image $\boldx^\mathrm{rec}$ as 
\begin{equation}\label{eq:dip}
	\bm{\theta}^\star = \argmin_{\bm{\theta}} \, L\left(\bm{F}_{\bm{\theta}} \left(\boldz\right), \boldx^{\mathrm{rec}}\right), \quad \boldx^{\mathrm{pp}} =  \bm{F}_{\bm{\theta}^\star} \left(\boldz\right) \, ,
\end{equation}
where $\boldz$ is a random image input. This approach relies on the implicit regularization imposed by the architecture of the \ac{NN} $\bm{F}_{\bm{\theta}}$ which prevents over-fitting, the structure of a \ac{NN} being sufficient to capture the low-level information of the image. This section describes the denoising methods in \ac{ET} that utilize \ac{DIP}.

A \ac{3D} U-Net was trained in \cite{cui2018ct,cui2019pet} to predict denoised \ac{PET} images using \ac{CT}/\ac{MRI} images as the high-quality prior input (i.e., $\boldz$ in \eqref{eq:dip}) and noisy \ac{PET} images as the label. It was observed that using a prior image  further improved the results rather than training the \ac{NN} on a random noise input (as done in \cite{ulyanov2018deep}). The network was trained on a single set of \ac{MR} images and noisy \ac{PET} volumes and then evaluated on simulated phantoms as well as real datasets. The loss function used was the \ac{MSE} combined with the \ac{LBFGS} algorithm \cite{liu1989limited}. The \ac{DIP} method was extended to dynamic \ac{PET} denoising in \cite{hashimoto2019dynamic}. The \ac{3D} U-Net in this work was trained on static \ac{PET} images as inputs to the network and noisy dynamic \ac{PET} images as training labels. DeepRED \cite{sun2021dynamic} is a method that utilized \ac{DIP} with \ac{RED} \cite{mataev2019deepred}. A U-Net-like architecture was trained on noisy dynamic \ac{PET} images with \ac{MSE}. The input to the network was a random noise vector. \citeauthor{hashimoto20214d} extended their previous work  \cite{hashimoto2019dynamic} in \cite{hashimoto20214d} where they proposed two modular approaches to  dynamic \ac{PET} image denoising. The first module consisting of a \ac{3D} U-Net extracted features from static \ac{PET} images. The extracted features were then fed to a reconstruction module consisting of a typical \ac{CNN} with convolution layers that predicted denoised images while being trained on noisy dynamic \ac{PET} volumes. Simultaneous denoising of dynamic \ac{PET} images was proposed by \citeauthor{yang2022simultaneous}~\cite{yang2022simultaneous}. Their network consisted of multiple convolutional layers, taking as input time averaged \ac{PET} images with the training labels being the noisy dynamic \ac{PET} volume. A variation of this network called \ac{DDIP} was also proposed which additionally generates the time averaged \ac{PET} images.

\subsubsection{Noise2noise}

\Ac{N2N}  is a \ac{SSL} technique, i.e., a machine learning technique that trains models using  input data only, without labeled data and explicit supervision. Labeling data takes considerable time and effort and in some situation obtaining gold standard-labeled data is impossible. \Ac{SSL} has the advantage of significantly increasing the number of datasets for model training as it does not require labeled data. In medical imaging techniques that employ  ionizing radiation, such as X-ray and gamma-ray, obtaining a clean target (label) with a high radiation dose can elevate the potential health risks associated with radiation exposure. Although it is possible to obtain a clean target by increasing the scanning time, it may cause image blurring or distortion by increasing the likelihood of patient movement. 

The \ac{N2N} approach, as proposed by \citeauthor{lehtinen2018noise2noise}~\cite{lehtinen2018noise2noise},  trains a deep \ac{NN} $\boldF_{\bm{\theta}}$ parametrized with $\bm{\theta}$ for image denoising using $K$ noisy image pairs, i.e., two noise instances of the same image, as input and target, as follows:
\begin{equation}
	\bm{\theta}^\star = \argmin_{\bm{\theta}} \, \sum_{k=1}^K\left\|\boldF_{\bm{\theta}}\left(\boldx^{\mathrm{clean}}_k+\bm{\epsilon}_{k, 1}\right)-\left(\boldx^{\mathrm{clean}}_k+\bm{\epsilon}_{k, 2}  \right)\right\|^2
\end{equation}
where $\bm{\epsilon}_{k,1},\bm{\epsilon}_{k,2}$ are independent noise realizations for each image $\boldx^{\mathrm{clean}}_k$, $k=1,\dots,K$. This \ac{SSL} technique, which only requires the input and target noise distribution to be identical and independent, has demonstrated its efficacy in reducing various types of noise, including Gaussian and Poisson noise.

There has been a growing interest in applying the \ac{N2N} to medical images, such as  \ac{CT} and \ac{MRI} \cite{wu2021low,fang2021iterative,jung2022mr}. In these studies, the denoising efficacy of \ac{N2N} was compared to that of conventional denoising techniques. \Ac{CT} images reconstructed using \ac{N2N} prior images yielded better \ac{RMSE} and \ac{SSIM}, as well as improved texture preservation as compared with conventional methods using total variation, \ac{NLM}, and convolutional sparse coding \cite{wu2021low}. \citeauthor{fang2021iterative}~\cite{fang2021iterative} showed that using a \ac{N2N}-denoised image as a prior within \ac{MBIR} showed promising results in suppressing noise while preserving subtle structures when applied to spectral \ac{CT} data for material decomposition. \citeauthor{jung2022mr}~\cite{jung2022mr} evaluated the performance of \ac{N2N} for image quality improvement in sub-millimeter resolution \ac{3D} \ac{MR} images. In this study, the K-space data of \ac{3D} \ac{MR} images were split into two separate sets with  independent noise realizations, which was achieved by undersampling data alternatively along the $k$-$z$-axis and estimating missing data using a GRAPPA kernel. Volumetric accuracy, as well as image quality, was improved by \ac{N2N}, which utilized only a single fully sampled K-space data.

\begin{figure*}
    \centering
    \includegraphics[width=0.9\linewidth]{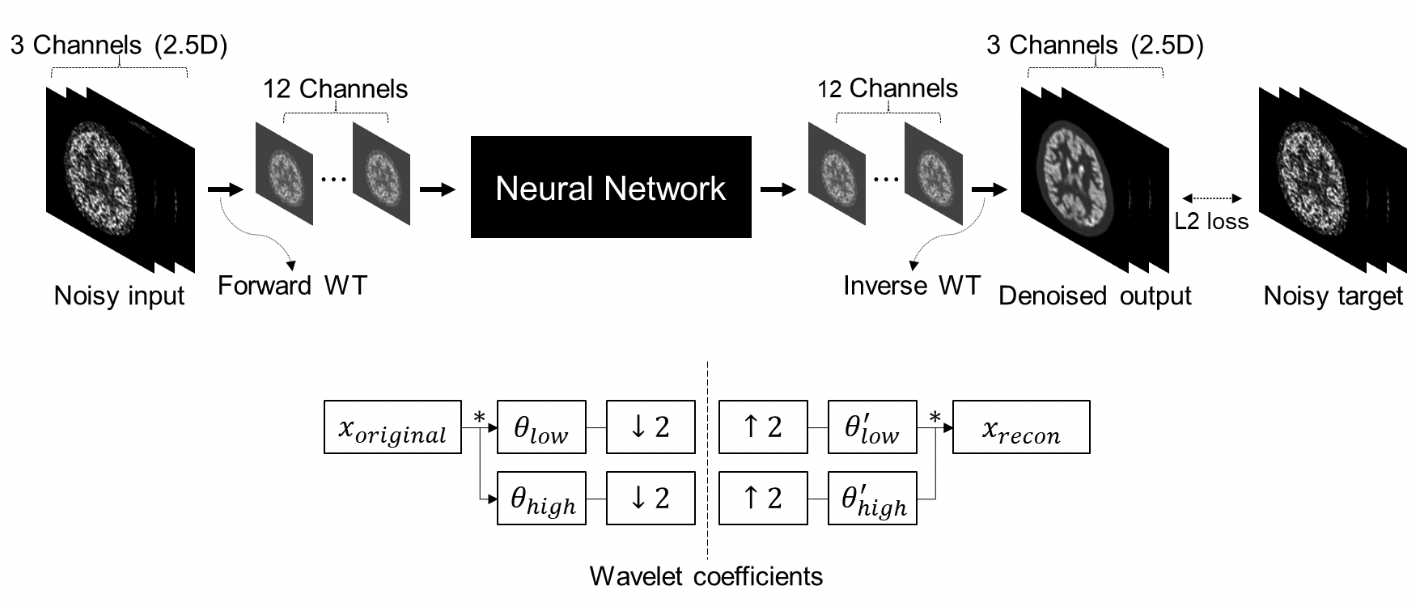}
    \caption{Schematic of \ac{N2N} network model improved by the incorporation of \acp{WT}. Reprint from \citeauthor{kang2021noise2noise}~\cite{kang2021noise2noise}.}
    \label{fig:n2n}
\end{figure*}

\begin{figure*}
    \centering
    \includegraphics[width=0.9\linewidth]{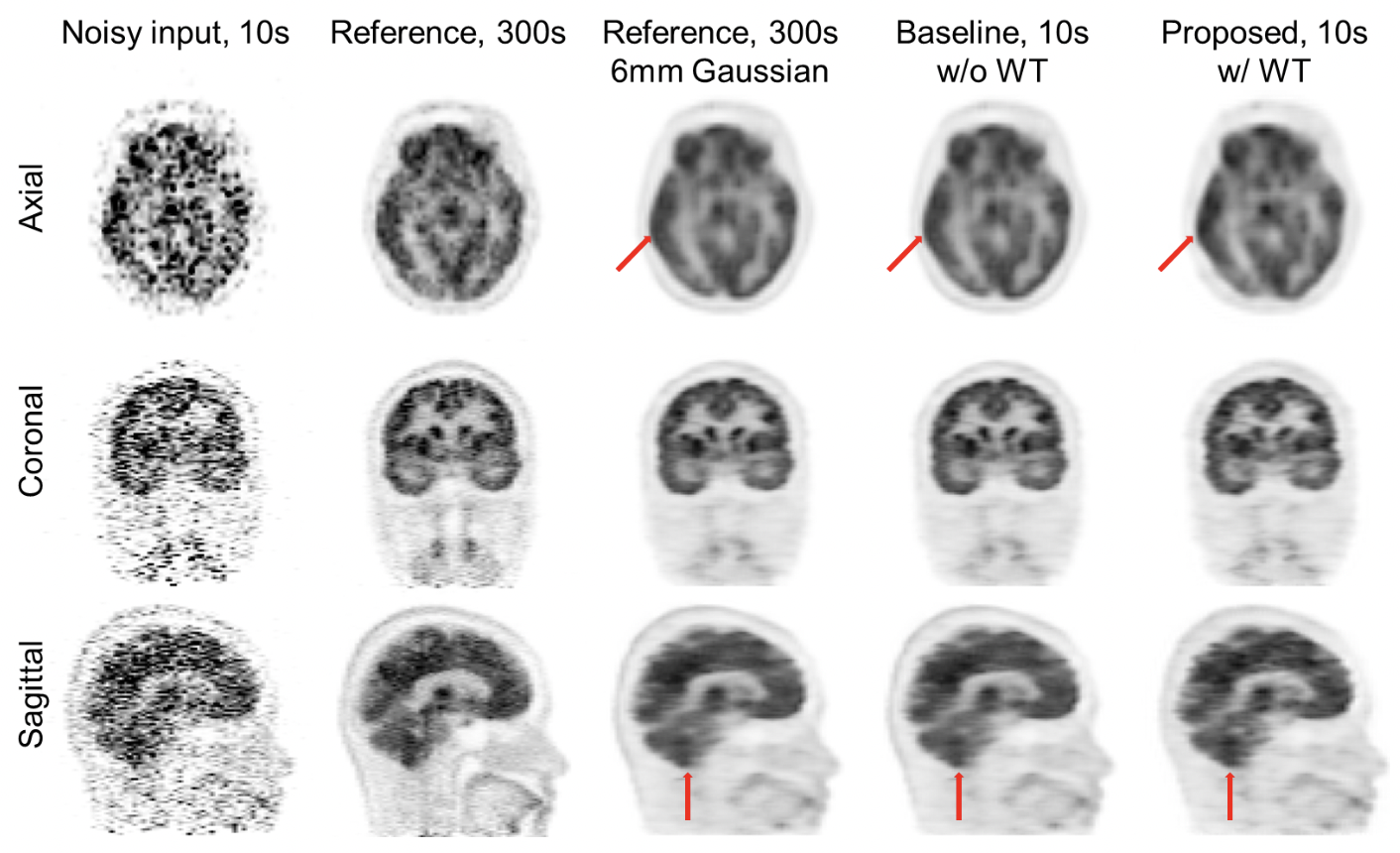}
    \caption{Full count image, noisy input, Gaussian filtered and denoised images using the N2N without and with incorporating the trainable \ac{WT} for clinical data. Reprint from \citeauthor{kang2021noise2noise}~\cite{kang2021noise2noise}.}
    \label{fig:full-count}
\end{figure*}

\Ac{N2N} and \ac{SSL} technology are also potentially useful for improving the image quality and diagnostic performance in  \ac{ET}. List-mode data recorded in nuclear medicine image scans and containing detail information about the location, energy, and time of each detected event allows for flexible data binning. Its flexibility has provided excellent platform for investigating and implementing \ac{N2N} technologies that require identical and independent dataset as input and target for network training. A \ac{N2N} application study on [\textsuperscript{18}F]FDG brain \ac{PET} utilized short time-bin images (10s--40s) generated from list-mode data and showed equivalent \ac{PSNR} of \ac{N2N} outcomes compared to supervised learning with 300s images as target, at all tested noise levels \cite{yie2020self}. The usefulness of \ac{N2N} for noise reduction in [\textsuperscript{15}O]water dynamic \ac{PET} and [\textsuperscript{99m}Tc]MDP/DPD \ac{WB} bone scan studies was also reported \cite{wu2020deep,yie2021self}. In addition, \citeauthor{chan2019noise}~\cite{chan2019noise} proposed a technique to improve the accuracy and robustness of \ac{N2N} network trained for \ac{WB} \ac{PET} image denoising by mitigating the high variance in \ac{N2N} denoising outcomes. The variability is mainly caused by the spatially non-stationary nature of \ac{PET} image noise distribution. In this study, instead of training the network with  pairs of individual noisy realizations, the number of training samples was increased by pairing a single noisy realization with an ensemble of noisy realizations at the same count level. When applied to low-count \ac{WB} \ac{PET} images, the original \ac{N2N} produced speckle and clustered noise artifacts. Nevertheless, the proposed method was effective in reducing the noise while preserving natural noise texture. An endeavor has been also undertaken to enhance the generalization ability of \ac{N2N}-based noise reduction algorithm through the incorporation of \acp{WT} into the \ac{N2N} network \cite{kang2021noise2noise}. The proposed method entails the utilization of the forward \ac{WT} to decompose a given noisy image into its low-pass and high-pass frequency components. These data are then fed into an \ac{N2N} network, followed by an inverse \ac{WT}. The final output was compared with another noisy image as in typical \ac{N2N} framework. The forward and backward \ac{WT} coefficients were also determined through training, thereby enabling the proposed method to outperform the original \ac{N2N} method in suppressing artifacts and preserving abnormal uptakes. This method is shown in Fig.~\ref{fig:n2n} and Fig.~\ref{fig:full-count}.

Multiple variants of \ac{N2N} have been developed to mitigate different types of noise and enhance the efficacy of self-supervised denoising algorithms. These variants include Noiser2Noise, Noise2Void, and Noise2Self \cite{moran2020noisier2noise,krull2019noise2void,batson2019noise2self}. When applied to low-count [\textsuperscript{18}F]FDG brain \ac{PET} images, Noiser2Noise, which requires only a single noisy realization of each training sample, was more effective than \ac{N2N} in preserving the noise texture of the input images \cite{yie2020self}. Noise2Void, another \ac{SSL} technique that does not require paired training samples, outperformed traditional denoising methods for \ac{PET} when pre-trained through transfer learning and guided by anatomical images \cite{song2021noise2void}.

\subsection{Multi-modality}\label{sec:multi}

In hybrid \ac{PET}/\ac{CT}, \ac{PET}/\ac{MR}, and \ac{SPECT}/\ac{CT} systems, the anatomical information derived from \ac{CT} and \ac{MRI} can also be used to enhance the denoising performance of \ac{PET} and \ac{SPECT}. Numerous studies using \acp{CNN} in brain \ac{PET}/\ac{MR} datasets achieved substantial dose reduction, sometimes up to 99\%, and incremental improvements in image quality and resolution have been demonstrated by combining \ac{PET} and \ac{MR} images as multiple channels into the network compared to using \ac{PET} images alone as input \cite{xiang2017deep,chen2019ultra,wang20183dauto,chen2020generalization,liu2019higher,chen2021true}. While \acp{CNN} requirement on training datasets is high, a relatively low-complexity \ac{CNN} (micro-net) that is more robust to very limited amounts of training data was proposed for \ac{MR}-guided \ac{PET} denoising, and demonstrated to have robust performance \cite{da2020micro}. Other networks structures, such as a spatial adaptive and transformer fusion network \cite{zhang2022spatial}, have also been proposed to improve \ac{PET} denoising by more effectively incorporating \ac{MRI} information. 

Not only helping noise reduction, anatomical \ac{CT} and/or \ac{MRI} information can also improve the \ac{PET} image resolution. For example, anatomically-guided \ac{PET} reconstruction using the Bowsher prior \cite{bowsher1996bayesian} can be generated by a \ac{CNN} in the image, thus space allowing the generation of anatomically-guided high-resolution \ac{PET} images without the need to access raw data and reconstruction console \cite{schramm2021approximating}. When \ac{PET} raw data are available, incorporating sinogram-based physics into the loss function of \ac{PET}/\ac{MR} networks has been shown to further improve the denoising performance that is more robust to \ac{OOD} data \cite{sudarshan2021towards}. 

In addition to the strategy of incorporating anatomical images as multi-channel inputs, such information can also be input to the network through a \ac{SA} mechanism, for unsupervised \ac{PET} denoising for example \cite{onishi2021anatomical,xie2021anatomically}. 

In another strategy, investigations using \ac{MR} and \ac{CT} images as prior information to guide \ac{DIP} (cf. Section~\ref{sec:dip}) have been performed for brain and body \ac{PET}/\ac{CT} datasets \cite{cui2018ct,cui2019pet,gong2019machine}. 
The group further extended the anatomical-guided \ac{DIP} approach to direct parametric reconstruction framework, where \ac{CT} and \ac{MR} images were incorporated as the network input to provide a manifold constraint, and also utilized to construct a kernel layer to perform non-local feature denoising \cite{gong2021direct}.

One requirement of incorporating anatomical information into \ac{PET} denoising is the integrated or simultaneous scanners such as \ac{PET}/\ac{MR}, which are still limited to widespread use. A study suggested that accurate full-dose amyloid \ac{PET} images can be generated from low-dose \ac{PET} and either simultaneous or non-simultaneous (acquired up to 42 days apart) \ac{MR} images, broadening the utility of low-dose amyloid \ac{PET} imaging \cite{chen2022investigating}. 

While a large number of hybrid \ac{PET} denoising work were performed for brain \ac{PET}/\ac{MR} datasets, investigations using \ac{CT} and/or \ac{MRI} information for denoising body \ac{PET} images were also performed. In an application to cardiac viability [\textsuperscript{18}F]FDG imaging in patients with ischemic heart disease, both attenuation correction \ac{CT} and low-count \ac{PET} were input into networks through two channels, leading to effective denoising \cite{ladefoged2021low}. In a study of [\textsuperscript{68}Ga]PSMA prostate \ac{PET}/\ac{MR}, 50\% dose reduction could be achieved by using a discrete-\ac{WT} \ac{CNN} with \ac{MRI} priors \cite{deng2022low}. Rather than concatenating the \ac{MR} and \ac{PET} at the input level, a study combined them in the feature space with attention-weighted loss, and applied the methods to \ac{WB} PET for children and young adults lymphoma patients \cite{wang2021artificial}. 

While brain \ac{MR} images are usually well registered with \ac{PET} images, for \ac{WB} applications, mismatch between \ac{PET} and \ac{CT}/\ac{MR} due to motion could complicate \ac{PET} denoising methods that incorporate anatomical information. A study showed that when \ac{CT} and \ac{PET} are aligned, incorporating \ac{CT} as additional channels improves the quantitative accuracy of lung lesions derived from denoised low-count \ac{PET} images. However, when \ac{CT} and \ac{PET} are misaligned, incorporating \ac{CT} information resulted in additional lesion quantification bias as compared with using \ac{PET} data only \cite{lu2019investigation}. The results suggest that motion correction and image registration are important pre-processing steps when incorporating anatomical information into \ac{PET} denoising.

Similar approaches described above can also be applied to hybrid \ac{SPECT}/\ac{CT} scanners. In a \ac{SPECT} bone imaging application, a lesion-attention weighted U$^2$-Net incorporating both 1/7-count \ac{SPECT} and resampled attenuation correction \ac{CT} can lead to the successful generation of synthetic standard-count \ac{SPECT} images \cite{pan2022ultra}.

\subsection{Diffusion Models}\label{sec:dm}

\Acp{DM} have arisen as an alternative to \acp{GAN} for image generation and other tasks \cite{dhariwal2021diffusion}, and have been widely explored in medical imaging \cite{kazerouni2023diffusion}. In the following paragraph we briefly summarize score-based \acp{DM} for image generation as described in \citeauthor{ho2020denoising}~\cite{ho2020denoising}.

Assume that the (clean) feasible images  $\boldx$ are distributed according to some unknown \ac{PDF} $p_0(\boldx)$ that we wish to use to randomly generate images. The forward diffusion is achieved by generating a collection of images $\left(\boldx_t\right)_{t=1,\dots,T}$, starting from a clean image $\boldx = \boldx_0$ drawn from $p_0$ (i.e., by randomly choosing an image from the training dataset), as a Markov chain defined as
\begin{equation}\label{eq:fdiff}
	\boldx_t \mid \boldx_{t-1}  \sim \mathcal{N}\left( \sqrt{\alpha_t} \boldx_{t-1}  , (1-\alpha_t) \bm{I}  \right)
\end{equation}
where $\bm{I}$ is the identity operator in the image space, and $\left(\alpha_t\right)_{t=1,\dots,T}$ is a collection of parameters in $]0,1]$ defined such that $\boldx_T$ is (approximately) a white noise. Sampling an image according to the initial distribution $p_0$ can therefore be achieved by reversing the diffusion process, i.e., starting from a white noise $\boldx_T$ from which $\boldx_{T-1}$ is generated, and so on until $\boldx_0$. Unfortunately, the reverse conditional \acp{PDF} $p\left(\boldx_{t-1} \mid \boldx_{t}\right)$ are unknown. Instead, the approximate model is used: 
\begin{equation}
	\boldx_{t-1} \mid \boldx_{t} \sim \mathcal{N} \left(  \boldmu(\boldx_t,t),\sigma^2(t)\bm{I} \right)
\end{equation} 
where $\sigma^2(t)$ is a known function of $\alpha_t$ and $\boldmu(\boldx_t,t) =  \frac{1}{\sqrt{\alpha_t}} \boldx_t +  \frac{1-\alpha_t}{\sqrt{\alpha_t}} \nabla \log p_t (\boldx_t) $ where $p_t$ is the \ac{PDF} of $\boldx_t$. The score function $ \bm{s}_t(\bm{x}_t) = \nabla \log p_t (\boldx_t)$ is untractable and is therefore replace by a \ac{NN} $\bm{S}_{\bm{\theta}}(\boldx_t,t)$, where the parameter $\bm{\theta}$ is trained unsupervisingly via score-matching from several instances of $\left(\boldx_t\right)_{t=1,\dots,T}$ generated following \eqref{eq:fdiff} and by sampling $\boldx_0$ from a training dataset of clean images \cite{vincent2011connection}.

In addition to generating images, \acp{DM} can also be used for inverse problems such as denoising and reconstruction. One type of approach, namely \acp{CDM}, consists in generating a noise-free  image $\boldx$ according to the a-posteriori \ac{PDF} $p_0(\cdot | \boldy)$, where $\boldy$ is the measurement (in our case, $\boldy = \boldx^{\mathrm{noisy}}$), by successive sampling of 	$\boldx_{t-1}  | \boldx_{t},\boldy$. This can be achieved using the same above-mentioned methodology using the conditional score $\bm{s}_t(\bm{x}_t,\boldy) = \nabla\log p_t (\cdot | \boldy )(\boldx_t)$. The conditional score can be approximated score-matching in two ways: (i) by conditional score-matching of a \ac{NN}  $\bm{S}_{\bm{\theta}}(\boldx_t,\boldy,t)$ that takes as input both the image $\boldx_t$ for all $t$ and the measurement  $\boldy$ \cite{song2020score,dhariwal2021diffusion,liu2023dolce} (supervised), and (ii)  unconditional score-matching of $\bm{S}_{\bm{\theta}}(\boldx_t,t)$ (cf. previous paragraph)  combined with the Bayes rule and  an approximation of the posterior distribution  $p(\boldy | \boldx_t)$ \cite{chung2022diffusion,kawar2022denoising} (unsupervised). Note that approach (ii) requires the knowledge of the noise model of $\boldy$. With a different formulation but in the same spirit, \citeauthor{mardani2023variational}~\cite{mardani2023variational} proposed to approximate the a-posteriori \ac{PDF} $p_0(\cdot | \boldy)$ by minimization of the Kullback-Leibler divergence between a standard Gaussian model and the posterior, which yields to a penalized least-squares optimization problem with a score-matching penalty. 

Recently,  \citeauthor{gong2023pet}~\cite{gong2023pet} used a \ac{CDM} for brain \ac{PET} image denoising with incorporation of the \ac{MR} image as prior information  $\boldy$ to the approximated conditional score $\bm{S}_{\bm{\theta}}(\boldx_t,\boldy,t)$ trained from a clean \ac{PET}/\ac{MRI} dataset. \citeauthor{pan2023full}~\cite{pan2023full} used a \ac{CDM} reinforced with a consistency model \cite{song2023consistency} to improve the efficiency. \citeauthor{shen2023pet}~\cite{shen2023pet} proposed to utilize diffusion models for low-count \ac{PET} image denoising through a bidirectional condition diffusion probabilistic model, which was validated on \ac{WB} \ac{PET} datasets. \citeauthor{jiang2023pet}~\cite{jiang2023pet} proposed an unsupervised \ac{PET} enhancement framework through the latent diffusion model, which can be trained only on standard-count \ac{PET} data. Instead of a Gaussian noise, a Poisson noise was inserted in the diffusion process to better accommodate \ac{PET} imaging. Also, a \ac{CT}-guided cross-attention was proposed to incorporate additional \ac{CT} images into the inverse process. \citeauthor{han2023contrastive}~\cite{han2023contrastive} proposes to generate high-quality \ac{PET} based on the diffusion models through a coarse-to-fine \ac{PET} reconstruction framework that consists of a coarse prediction module and an iterative refinement module. 

Apart from \ac{PET} image denoising, \citeauthor{xie2022brain}~\cite{xie2022brain} proposed to generate synthetic \ac{PET} image from \ac{MR} images based on the diffusion models. \citeauthor{singh2023score}~\cite{singh2023score} proposed a diffusion model-based \ac{PET} image reconstruction framework, where the \ac{PET} forward model (eq. \eqref{eq:expectation} and eq. \eqref{eq:poisson}) was utilized together with the score function (i.e., approach (ii) above with the \ac{PET} raw data $\boldy$) for high-quality \ac{PET} image generation. Their method was further improved by adding \ac{MR} anatomical prior. 

Despite their potential in image denoising for \ac{ET}, it is important to note that \acp{DM} are relatively novel in this field. As of the current literature review, there is a limited number of research papers that have explored the application of \acp{DM} to denoise (or reconstruct) \ac{ET} images. This scarcity of studies underlines the emerging nature of \acp{DM} in \ac{ET} image processing, and as a result, it represents an exciting and promising avenue for further research and development.

\section{Discussion and Conclusion}\label{ref:disc}

\begin{table*}
	\centering
	\caption{Summary of the reviewed articles}\label{table:summary}
	\scriptsize
	\begin{tabular}{llllll}
		\hline
		Application    & Supervised & Method/Architecture  & Anatomical Information & References   \\
		\hline 
		\textbf{Section~\ref{sec:supervised}}: Supervised Methods& & & & \\
		Brain  \ac{PET} & yes &  \ac{CNN}, U-Net & no & \cite{xu2017200x,spuhler2020full} \\
		Brain \& Chest  \ac{PET} & yes &  \ac{ResNet} & no & \cite{gong2018pet} \\
		Brain  \ac{PET}  & yes  & \ac{DAE} & no&\cite{klyuzhin2019use} \\
		Chest  \ac{PET} & yes &  U-Net & no&\cite{schaefferkoetter2020convolutional} \\
		Whole-body   \ac{PET} & yes &  U-Net & no&\cite{liu2022personalized} \\
		\ac{MP} \ac{SPECT}& yes &  \ac{CNN}, U-Net, \ac{ResNet}  & no&\cite{ramon2020improving,reymann2019u,shiri2020standard,liu2021accounting} \\
		Chest  \ac{PET} & yes & \ac{GAN}, \ac{cGAN}, \ac{CycleGAN}, \ac{WGAN} &no& \cite{zhou2020supervised,gong2020parameter,geng2021content,xue2021cross,zhou2021mdpet,xue2021lcpr} \\
		Chest  \ac{PET} & yes & \ac{CNN}, U-Net, \ac{GAN}    & \ac{CT} &\cite{lu2019investigation} \\	
		Brain \ac{PET} & yes  & \ac{GAN} &no& \cite{ouyang2019ultra,jeong2021restoration} \\
		Brain  \ac{PET} & yes &  \ac{GAN}, \ac{cGAN},\ac{CycleGAN}  &no& \cite{wang20183d,zhao2020study,du2020iterative} \\
		Brain  \ac{PET} & yes &  \ac{cGAN}  &\ac{MRI}& \cite{wang20183dauto} \\
		Whole-body \ac{PET} & yes &  \ac{CycleGAN} &no& \cite{lei2019whole,sanaat2021deep,hu2020dpir} \\
		\ac{MP} \ac{SPECT}& yes & \ac{cGAN}, attention-based \ac{GAN} &no&  \cite{sun2022dual,sun2022deep,sun2022pix2pix,sun2023fast,aghakhan2022deep} \\
		\hline 
		\textbf{Section~\ref{sec:self-atten}}: \Ac{SA} Mechanisms&  & & &\\
		Whole-body  \ac{PET} & yes &  U-Net + \ac{SA} block & no & \cite{xue20203d} \\
		Chest  \ac{PET} & yes &  U-Net + \ac{SA} block & no & \cite{xie2020generative} \\
		Chest \ac{PET} & yes &  U-Net + \ac{SA} block & \ac{CT}, \ac{MRI}&\cite{xie2021anatomically} \\
		Brain  \ac{PET} & yes &  \ac{GAN} + \ac{SA} block & no&\cite{fu2023aigan} \\
		\ac{WB} \ac{PET} & yes &  \ac{CNN}, \ac{CycleGAN} + \ac{SA} block & no&\cite{shang2022short,li2023deep} \\
		Brain  \ac{PET} & no & \ac{CNN}  + \ac{SA} block  & \ac{MRI} &\cite{onishi2021anatomical} \\		
		\ac{WB}  \ac{PET}  & yes & Transformer & no&\cite{luo20213d,jang2022spach} \\
		Brain,  \ac{WB} \ac{PET}& yes & U-Net, \ac{GAN}, Transformer & \ac{MRI}& \cite{zhang2022spatial,wang2023low} \\
		Brain \ac{PET}  & yes & Transformer & no&\cite{hu2022transem} \\
		\hline 
		\textbf{Section~\ref{sec:unsup}}: Unsupervised Methods & & & & \\
		\ac{WB}  \ac{PET} & no & \ac{DIP} & \ac{CT}, \ac{MRI}& \cite{cui2018ct,cui2019pet} \\
		Brain  \ac{PET} & no & \ac{DIP} & no & \cite{hashimoto2019dynamic,sun2021dynamic,hashimoto20214d,yang2022simultaneous} \\
		Brain  \ac{PET} & no & \ac{N2N} & no & \cite{yie2020self,kang2021noise2noise,song2021noise2void} \\
		Bone \ac{SPECT} & no & \ac{N2N} & no & \cite{yie2021self} \\
		$^{15}$O water \ac{PET}  & no & \ac{N2N} & no & \cite{wu2020deep} \\
		\ac{WB}  \ac{PET} & no & \ac{N2N} & no & \cite{chan2019noise} \\
		\hline 
		\textbf{Section~\ref{sec:multi}}: Multi-modality& & & & \\
		Brain  \ac{PET} & yes &  \ac{CNN}, U-Net & \ac{MRI} & \cite{xiang2017deep,chen2019ultra,chen2020generalization,liu2019higher,chen2021true,da2020micro,schramm2021approximating,sudarshan2021towards,chen2022investigating} \\
		Brain  \ac{PET} & no & \ac{DIP} & \ac{MRI} & \cite{gong2021direct} \\
		$^{68}$Ga  \ac{PET} (prostate) & yes & \ac{CNN}    & \ac{MRI}&\cite{deng2022low} \\		
		\ac{MP} \ac{PET} & yes & U-Net  & \ac{CT} &\cite{ladefoged2021low} \\	
		\ac{WB}  \ac{PET} & yes & \ac{CNN}  & \ac{MRI} &\cite{wang2021artificial} \\	
		Bone \ac{SPECT} & yes & U-Net  & \ac{CT} &\cite{pan2022ultra} 	\\
		\hline 
		\textbf{Section~\ref{sec:dm}}: \Acp{DM} & & & &\\
		Brain  \ac{PET} & no & \ac{DM} & \ac{MRI} & \cite{gong2023pet} \\
		\ac{WB}  \ac{PET} & no & \ac{DM} & no & \cite{pan2023full,shen2023pet} \\
		\ac{WB}  \ac{PET} & no & \ac{DM} & \ac{CT} & \cite{jiang2023pet} \\
		Brain \ac{PET} & yes & \ac{DM} & no & \cite{han2023contrastive} \\
		Brain \ac{PET} (image synthesis) & yes & \ac{DM} & \ac{MRI} & \cite{xie2022brain} \\
		Brain \ac{PET} (image reconstruction) & no & \ac{DM} & \ac{MRI} & \cite{han2023contrastive} \\
		\hline
	\end{tabular}
\end{table*}

Reduction in the dosage of the radio-pharmaceutical administered to a patient undergoing a functional imaging scan is essential to reduce the risk and radiation the patient is being exposed to. Low-dose imaging aims to produce images with quality on par with regular-dose imaging while operating on a fraction of the radio-pharmaceutical. Getting a clear image in \ac{ET} imaging is a difficult task due to attenuation, scatter, and the ill-posed nature of the physics model used to characterize \ac{PET} and \ac{SPECT}. Low-dose \ac{ET} further complicates an already challenging problem. Images are produced by converting the detector data into readable images through the process of reconstruction. Low-dose \ac{ET} reconstructed images suffer from noise. One way to tackle this noise is through the post-processing of the reconstructed images. 

Image processing methods like Gaussian filtering have been implemented over the years to tackle the reconstructed noisy images. Recently deep learning-based methods have been found to surpass traditional algorithms in imaging tasks. In this article, various post-processing approaches that utilize deep \acp{NN} have been discussed at length. These approaches have been appropriately classified into various sub-sections to facilitate better readability and to distinguish them from one another. U-Net was the earliest \ac{NN} adapted for denoising. Adding a discriminator network to a U-Net resulted in \acp{GAN}. These networks were discussed in the Section~\ref{sec:supervised}. Recently developed supervised methods include a self-attention mechanism to U-Net and \acp{GAN} were highlighted in Section~\ref{sec:self-atten}. Deep image prior and \ac{N2N} methods are the most popular unsupervised \ac{NN} approaches, and were discussed in Section~\ref{sec:unsup}.  Section~\ref{sec:multi} presents methods that utilize multi-modality input data for training \acp{NN}  Finally, Section~\ref{sec:dm} covers recent trends in \acp{DM}. The primary focus of this article has been to discuss the diverse set of \ac{NN}-based approaches while keeping up with the latest trends in the denoising regime. The presented papers are summarized in Table~\ref{table:summary}.

One of the main challenges in training a \ac{NN} is the availability of data. There is an increasing number of publicly available datasets such as Brainweb  \cite{cocosco1997brainweb}. However, these are limited to the use cases they have been compiled for. There is a concern about using \acp{NN} trained on one specific dataset to be generalized to a larger data pool. This challenge of dealing with \ac{OOD} data is addressed to an extent by unsupervised methods, which do not require large labeled datasets. However, unsupervised methods are yet to perform convincingly better than supervised methods. Efforts are being made in this regard to develop \acp{NN} that produce robust results independent of labeled data. 

Evaluation of the denoised images predicted is another aspect that all \ac{NN} methods need to be held accountable for. Typically, \ac{MSE} and \ac{SSIM} are used for quantitative comparison and analysis. Plot profiles and region of interest analysis for tumors are also utilized for a more thorough evaluation. Some articles have further employed a scoring system used by radiologists for the qualitative assessment of the images. Such efforts to check the images produced by \ac{NN}-based approaches are essential to build trust and translate these methods to clinical cases. 

\Ac{NN}-based denoising has indeed revolutionized low-dose \ac{ET} imaging. This article through the emphasis on both \ac{PET} and \ac{SPECT} has highlighted the plethora of ways deep learning has improved image quality through post-processing. We discussed the earliest methods as well as the most promising methods in recent times that have the potential to be translated into clinical usage. The advantage of denoising is that methods applicable to one modality can easily be implemented for other modalities also, and this review on low-dose \ac{ET} could be useful to image denoising and enhancement in other modalities too.

The necessity to compare the multitude of methods presented in this review poses a formidable challenge, primarily due to their application-specific nature, spanning cardiology, neurology, and oncology, as well as the utilization of diverse training datasets. In an ideal scenario, method performances would be evaluated using dedicated benchmark datasets specifically designed for comparative purposes, accompanied by standardized metrics akin to those employed in renowned challenges within the field. Drawing inspiration from successful model evaluation frameworks, such as the Ultra-low Dose PET Imaging Challenge \cite{lowdose-challenge}, it is paramount for the community to consider the creation of specialized datasets and standardized metrics aimed at comprehensively evaluating these methods. Such initiatives hold the potential to significantly enhance our ability to objectively compare and contrast the efficacy of different techniques across diverse medical applications, thus enabling more informed decisions and driving progress within the field.

\section*{Acknowledgment}

All authors declare that they have no known conflicts of interest in terms of competing financial interests or personal relationships that could have an influence or are relevant to the work reported in this paper. 

\AtNextBibliography{\footnotesize} 
\printbibliography

\end{document}